\documentclass[11pt]{article}
\usepackage{geometry}                
\usepackage[parfill]{parskip}    
\usepackage{graphicx}
\usepackage{amsmath, amsthm, amsfonts}
\usepackage{amssymb}

\usepackage{color}
     \usepackage[colorlinks]{hyperref}
\hypersetup{linkcolor=blue,%
citecolor=red,%
urlcolor=cyan}
\usepackage{url} 

\usepackage{epstopdf}
\begin{document}

\title{
Unusual isospectral factorizations of shape invariant \\ 
Hamiltonians with Scarf II  potential}

 \author{ Yi\u{g}it Can Acar$^a$\footnote{ Yigit.Can.Acar@ankara.edu.tr, ORCID: \href{http://orcid.org/0009-0006-3656-343X}{0009-0006-3656-343X}}, 
Lorena Acevedo$^b$\footnote{cindylorena.acevedo@uva.es, ORCID: \href{http://orcid.org/0000-0002-3110-6487}{0000-0002-3110-6487}} \, and\, \c Seng\"ul Kuru$^a$\footnote{kuru@science.ankara.edu.tr (Corresponding author), ORCID: \href{http://orcid.org/0000-0001-6380-280X}{0000-0001-6380-280X}}
\\ 
$^a$Department of Physics, Faculty of Science, Ankara University, 06100 Ankara, Türkiye\\
$^b$Departamento de F\'isica Te\'orica, At\'omica y Optica, Universidad de Valladolid,\\ 47071 
Valladolid, Spain}

\maketitle
	
\begin{abstract}
In this paper, we  search  the factorizations of the shape invariant Hamiltonians with Scarf II potential.
We find two classes; one of them is the standard real factorization which  leads us to a real hierarchy of potentials and their energy levels;
the other one  is complex and it leads us naturally to a hierarchy of complex Hamiltonians. 
We will show some properties of these complex Hamiltonians: they  are not parity-time (or PT)  symmetric, but their spectrum is real and isospectral to the Scarf II real Hamiltonian hierarchy. The algebras for real and complex shift operators (also called potential algebras) are computed; they consist of $su(1,1)$ for each  of them and the total potential algebra including both hierarchies is the direct sum $su(1,1)\oplus su(1,1)$.
\end{abstract}

\section{Introduction: Scarf II potential}
Scarf II potential (or Gendenshtein potential) \cite{CKS,Natanson}  was proposed to describe the atomic and molecular interactions (diatomic molecule potential) in quantum mechanics \cite{Eyube}. It is one of the shape invariant (SI) potentials  given for instance in \cite{CKS} with the following expression
\begin{equation}\label{v}
V(x)=\frac{B^2-(A+\frac{\gamma}{2})^2+\frac{\gamma^2}{4}+2 B (A +\frac{\gamma}{2}){\rm sinh}{\gamma x}}{({\rm cosh}\gamma x)^2}
\end{equation}
where $A$, $B$ and $\gamma$, in principle, are assumed to be real  parameters.

In the literature,  Scarf II potential has recently  attracted much attention, in general not by itself, but for its complexification  as a particular simple model to display analytically some  properties of complex potentials (see for instance \cite{Quesne,Levai,Bagchi,Ahmed,Pal,Kapoor22}).  

Complex potentials have interesting properties from the point of view of non-Hermitian Hamiltonians. One of them is the existence of real or complex spectrum depending on the existence of parity-time (PT) symmetry \cite{Bender98,Mostafazadeh,Correa} and whether this symmetry is spontaneously broken or not \cite{CKS,Ahmed01,Gangopadhyaya,Abhinav,Junker}.
In \cite{Quesne},  new non-PT-symmetric Hamiltonians were obtained including the Hamiltonian with Scarf II  potential by using group theoretical methods. Levai and Znojil \cite{Levai} studied the PT-symmetric Scarf II potential in order to see the relation of PT symmetry and supersymmetry (SUSY). Quesne and Bagchi  revisited the PT-symmetric Scarf II potential in \cite{Bagchi} and they computed  the bound-state wavefunctions together with their energy levels. They found also  new completely solvable rationally extended partners of complexified Scarf II potential.   Ahmed et al. in \cite{Ahmed}  studied in detail the crossing of spectrum levels of the complex PT-symmetric Scarf II potential. 

{In general, complex potentials (non-Hermitian Hamiltonians) emerges frequently in optical systems \cite{Gupta20,Miri13}. For example, the Scarf-II potential, or its complexifications, is used as the refractive index profile in \cite{Kapoor22}  and the broken and unbroken PT and SUSY potentials in optical systems related  with Scarf II potential were investigated in \cite{Pal,Kapoor22}. For complex potentials, PT symmetry may assure real eigenvalues, so it is an important important concept. In \cite{Abhinav}, the relation between SUSY quantum mechanics and PT-symmetry is considered in full detail and in this context a large class of PT or not PT-symmetric complex potentials is studied.  Non-Hermitian optical waveguides are considered in \cite{Principe15} and the scattering properties of complex potentials also discussed in \cite{Miri19,Mostafazadeh11}.}

This work, is focused just on the real Scarf II potential, not on its complexifications. We will see that this real Hamiltonian has  not only a real   hierarchy of potentials obtained by factorization. There is also a second complex hierarchy with a broken supersymmetry. We remark that the existence of simultaneous real and the complex hierarchies is not due to any complexification, simply, for the real Scarf II potential there exist these two compatible factorization hierarchies. {Remark that the real and complex Scarf II Hamiltonians of these two factorizations are not PT invariant.}
 
We show that  each of these hierarchies, real and complex, is linked to the Lie algebra $su(1,1)$. The corresponding factorization is associated to the parameters, $\beta$ or $\alpha$ of  Scarf II potential, respectively; while the real factorization changes $\beta $ in real units, the complex one changes the parameter $\alpha$ in  imaginary (complex) units. Besides, we show that the real shift operators  and the complex shift operators  commute and therefore the total shift (or potential) algebra for this problem is $su(1,1)\oplus su(1,1)$. 
This double real-complex factorization of Scarf II potential shown here is unique in the list of shape invariant potentials, as given for instance in  \cite{CKS}. The rest of SI potentials leading to double factorizations (for instance, trigonometric or hyperbolic P\"oschl-Teller potentials) have two simultaneous real factorizations. Therefore,  Scarf II is a very special case in the class of SI potentials and deserves to be examined in closer detail. The organization of this work is as follows. In section 2 we describe how such real and complex factorizations appear, as well as the spectrum of the  potentials in each hierarchy. The following section supply a discussion on the Lie algebras involved, to  arrive to  $su(1,1)\oplus su(1,1)$ as the algebra of the whole real-complex hierarchy.

\section{Factorizations}
Let us take the Scarf II Hamiltonian choosing  $\gamma=2$ in (\ref{v}), for the sake of simplicity, in the following form
\begin{equation}\label{hv}
H_{\alpha,\beta}(x)=-\frac{d^2}{dx^2}+V_{\alpha,\beta}(x),\qquad
V_{\alpha,\beta}(x)=\frac{{\alpha}^2-{\beta}^2+1+2 \alpha\beta\, {\rm sinh}{2x}}{({\rm cosh}2x)^2}
\end{equation}
Our aim is to factorize this Hamiltonian  in terms of real first order differential operators $A^\pm$ plus an extra constant as follows:
\begin{equation}\label{h1}
H_{\alpha,\beta}(x)=A^+ A^- +\mu
\end{equation}
The constant $\mu$ is called factorization energy and usually corresponds to ground state energy of the system.
Therefore, we will start by computing a real factorization and see how an additional complex one will automatically appear.
In the next sections we will characterize each factorization as well as their relations. 
\subsection{Real factorizations}

Suppose that both $\alpha$ and $\beta$ are real. 
The first order diferential operators $A^{\pm}$, are sometimes called shift operators, because they relate (or intertwine) Hamiltonians with  shifted parameters of the potential. The Hamiltonian can be writen as:
\begin{equation}\label{h11}
H_{\alpha,\beta}=A_{\alpha,\beta}^+ A_{\alpha,\beta}^- +\mu_{\beta}
\end{equation}
where
\begin{equation}\label{apm1}
A_{\alpha,\beta}^{\pm}=\mp\frac{d}{dx}+\frac{\alpha}{{\rm cosh}2x}+(\beta-1) {\rm tanh}2x, \qquad \mu_{\beta}=-(\beta-1)^2
\end{equation}
It can be easily seen from (\ref{apm1}) that the operators $A_{\alpha,\beta}^{\pm}$ are adjoint each other: $(A_{\alpha,\beta}^{+})^{\dag}=A_{\alpha,\beta}^{-}$.

Since the potential $V_{\alpha,\beta}(x)$ has the reflection symmetry $\alpha\to\tilde \alpha= -\alpha;\  \beta \to \tilde \beta= -\beta$,
then,  we have two types of sign election for the parameters giving rise to two  different but related factorizations.  By applying this symmetry to the factorization, (\ref{h11}) and operators (\ref{apm1}), we get another set of first order differential operators $\tilde{A}_{\alpha,\beta}^{\pm}$ that coincide with $A_{\alpha,\beta+2}^\mp$, together with a new constant $\tilde \mu_{\beta}=\mu_{\beta+2}$. So, the Hamiltonian $H_{\alpha,\beta}$ can also be written as:
\begin{equation}\label{h12b}
H_{\alpha,\beta}=\tilde{A}_{\alpha,\beta}^+\tilde{A}_{\alpha,\beta}^- +\tilde{\mu}_{\beta}=A_{\alpha,\beta+2}^- A_{\alpha,\beta+2}^+ +\mu_{\beta+2}
\end{equation}
where
\begin{equation}\label{apm2}
\tilde{A}_{\alpha,\beta}^{\pm}=\mp\frac{d}{dx}-\frac{\alpha}{{\rm cosh}2x}+(-\beta-1) {\rm tanh}2x, \qquad \tilde{\mu}_{\beta}=-(-\beta-1)^2
\end{equation}
From (\ref{h11}) and (\ref{h12b}),  the following factorizations and intertwining relations are found:
\begin{equation}\label{h12h}
H_{\alpha,\beta}=A_{\alpha,\beta}^+A_{\alpha,\beta}^- +{\mu}_{\beta}=A_{\alpha,\beta+2}^- A_{\alpha,\beta+2}^+ +\mu_{\beta+2}
\end{equation}
\begin{equation}\label{int1}
A_{\alpha,\beta}^- H_{\alpha,\beta}=H_{\alpha,\beta-2}A_{\alpha,\beta}^- ,\qquad A_{\alpha,\beta}^+ H_{\alpha,\beta-2}=H_{\alpha,\beta}A_{\alpha,\beta}^+
\end{equation}
Thus, we have obtained a real hierarchy of Scarf II Hamiltonians
$
H_{\alpha,\beta +2m}\,,\,\, m\in \mathbb Z
$. 

The ground state for $H_{\alpha,\beta}$ is given by
\begin{equation}\label{psi0}
A_{\alpha,\beta}^- \psi_{\alpha,\beta}^0=0,\qquad  
\psi_{\alpha,\beta}^0(x)=N_0\,e^{-\alpha \arctan(\tanh x)} (\cosh{2 x})^{(-\beta+1)/2}\,,\quad \beta>1
\end{equation}
where $N_0$ is normalization constant depending on the values of  $\alpha$, $\beta$ and the  upper index of $\psi_{\alpha,\beta}^0$ 
denotes the ground state.
Keeping in mind that the Gudermannian ${\rm gd}\, x$ is defined by \cite{Beyer}
\[
{\rm gd}\, x= 2 \arctan\Big(\tanh \frac1{2}x\Big),\qquad 
{\rm gd}' x={\rm sech} x
\]
then, the ground state wavefunction can be re-expressed as
\[
\psi_{\alpha,\beta}^0(x)=N_0\,e^{-\frac{\alpha}2\, {\rm gd}\, 2x} 
(\cosh{2 x})^{(-\beta+1)/2}\,,\quad \beta>1
\]
The parameter $\beta$ must satisfy $\beta>1$, in order  $\psi_{\alpha,\beta}^0$ be square integrable. The factorization energy $\mu_\beta$ corresponds to the ground state energy: $E_{\alpha,\beta}^0=\mu_{\beta}=-(\beta-1)^2$. 

According to the intertwining (\ref{int1}), the action of the real shift operators on the $n$th-excited state $\psi_{\alpha,\beta}^n $ of $H_{\alpha,\beta}$ is given by:
\begin{equation}\label{apmpsin}
A_{\alpha,\beta+2}^+\psi_{\alpha,\beta}^n \propto \psi_{\alpha,\beta+2}^{n+1},\qquad A_{\alpha,\beta}^-\psi_{\alpha,\beta}^n \propto \psi_{\alpha,\beta-2}^{n-1}\,
\end{equation}
where $n$ denotes the excitation level of the state and $n=0,1,2,\dots$\,. Then, any excited state can be obtained by the iterative action of the kind of operators ${A}_{\alpha,\beta}^{+}$ on the ground state eigenfunctions:
\begin{equation}\label{psi0n}
  A_{\alpha,\beta}^+\dots A_{\alpha,\beta-2n+2}^+ \psi_{\alpha,\beta-2n}^0\propto \psi_{\alpha,\beta}^n 
,\qquad E_{\alpha,\beta}^n = E_{\alpha,\beta-2n}^0
\end{equation}
under the condition $\beta-2n-1>0$.
The $n$th excited state solution in terms of  Jacobi polynomials for the real case can be found, for instance in \cite{CKS,Alvarez}. It has the form:
\begin{equation}\label{psin}
\psi_{\alpha,\beta}^n(x)=N_n\,i^n (1+y^2)^{-s/2} e^{-\lambda \tan^{-1}y} P_n^{(i\lambda-s-1/2,-i\lambda-s-1/2)}(iy),\qquad y=\sinh2x
\end{equation}
where $N_n$ is normalization constant, $P_n^{(\mu,\nu)}$ is for a Jacobi polynomial and $s=\frac{\beta-1}{2},\, \lambda=\frac{\alpha}{2}$  depend on the potential parameters \cite{Abramowitz}.

The corresponding energies and the conditions for bound states on the $\beta$-parameter are given by
\begin{equation}\label{e1}
E_{\alpha,\beta}^n=-(\beta-1-2n)^2,\quad n=0, 1,\dots,n_{\rm max}, \quad n_{max}=[\frac{\beta-1}{2}]\ \  {\rm and} \quad \beta>1,\ \beta-2n_{\rm max}-1>0 
\end{equation}
\begin{figure}[ht]
  \centering
\includegraphics[width=0.4\textwidth]{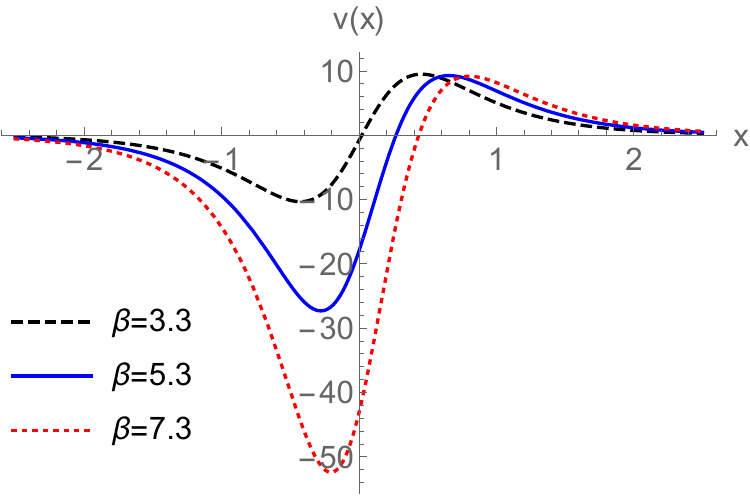}%
\hspace{1.5cm}
  \includegraphics[width=0.4\textwidth]{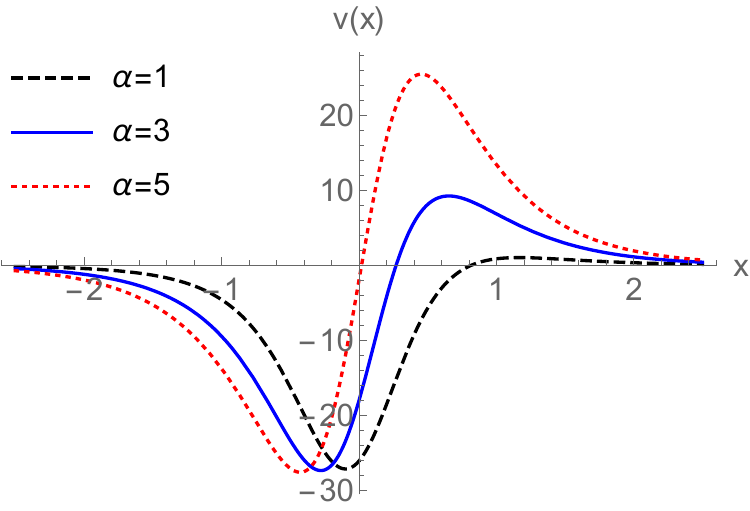}
\caption{\small
Plot of the real Scarf II potential for different values of parameters:  $\alpha=3, \beta_1=3.3, \beta_2=5.3, \beta_3=7.3 $ (left) and for  $\beta=5.3, \alpha_1=1, \alpha_2=3,  \alpha_3=5 $  (right).
}
\label{figuras1}
\end{figure}

\begin{figure}[ht]
  \centering
\includegraphics[width=0.5\textwidth]{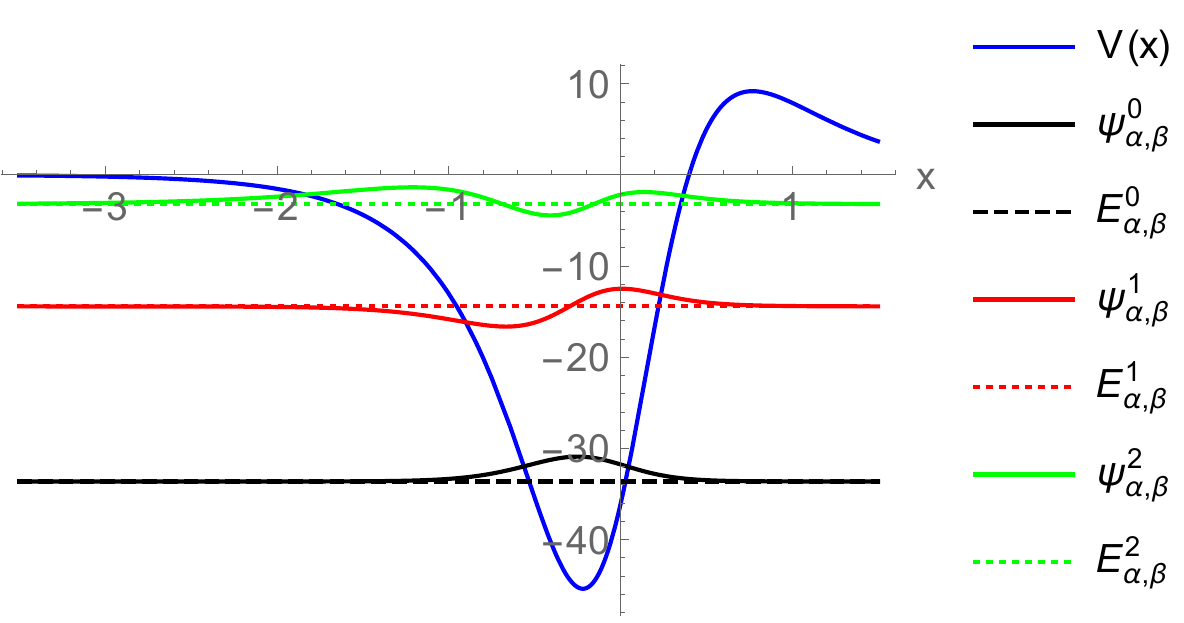}
\caption{\small
Plot of the energies, the eigenfunctions of  the real Scarf II potential for  $\alpha=3, \beta=6.8 $  and  $n=0,1,2 $ where $n_{max}=[(6.8-1)/2]=[2.9]=2$.
}
  \label{figuras2}
\end{figure}

{From Fig.~\ref{figuras1} (left) it can be appreciated that the parameter $\beta$ determines the depth of the potential and so the number of the bound states. However, the effect of $\alpha$, Fig.~\ref{figuras1} (right), seems to have no significative influence on the depth and therefore it will affect the shape  of eigenfunctions but not the spectrum (in agreement with (\ref{e1})).} In Fig.~\ref{figuras2}, we have shown the spectrum of  real Scarf II potential for some fixed values of the parameters. 

Remark that due to the symmetry $\alpha \to \tilde\alpha= - \alpha$, $\beta \to
\tilde\beta= - \beta$,
 if $ \beta>1$ then $\tilde \beta<-1$. Thus, we can define the ``left'' ground states 
 for $H_{\tilde\alpha,\tilde\beta}$  by
\begin{equation}\label{psi0b}
A_{\tilde\alpha,\tilde\beta+2}^+ \tilde\psi_{\tilde\alpha,\tilde\beta}^0=0,\qquad  
\tilde \psi_{\tilde\alpha,\tilde\beta}^0(x)=N_0\,e^{\frac{\tilde\alpha}2\, {\rm gd}\, 2x} (\cosh{2 x})^{(\tilde\beta+1)/2}
\end{equation}
Square integrability of the ground state $\tilde\psi_{\tilde\alpha,\tilde\beta}^0$, makes necessary that $\tilde \beta<-1$. The factorization energy corresponding to the ground state energy is $\tilde E_{\tilde\alpha,\tilde\beta}^0=\mu_{\tilde\beta+2}=-(\tilde\beta+1)^2$. 
The excited states can be obtained by the action of ${A}_{\tilde\alpha,\tilde\beta}^{-}$ on the ground state eigenfunctions:
\begin{equation}\label{psi0nb}
 A_{\tilde\alpha,\tilde\beta+2}^-\dots A_{\tilde\alpha,\tilde\beta+2n}^- \tilde\psi_{\tilde\alpha,\tilde\beta+2n}^0\propto \tilde\psi_{\tilde\alpha,\tilde\beta}^n \,,\qquad \tilde\beta+2n<-1
\end{equation}
\[
\tilde E_{\tilde\alpha,\tilde\beta}^n = \tilde E_{\tilde\alpha,\tilde\beta+2n}^0
=-(\tilde\beta+2n+1)^2
\]
Therefore, we have a complete symmetry between the hierarchy $\alpha,\beta> 1$ and
$\tilde \alpha = -\alpha, \tilde \beta = -\beta$:
\begin{equation}
 E_{\alpha,\beta}^0 = \tilde E_{\tilde \alpha,\tilde\beta}^0,\quad
  E_{\alpha,\beta}^n =\tilde E_{\tilde\alpha,\tilde\beta}^n
\end{equation}
The symmetry is implemented to the eigenstates:
\begin{equation}\label{psi0nc}
\tilde\psi_{\tilde\alpha,\tilde\beta}^n = \psi_{\alpha,\beta}^n\,
\qquad \beta - 2n>1,\quad  \tilde \beta +2n  <-1
\end{equation}
In summary, bound states will be present for $\beta>1$ or for $\tilde \beta<-1$.
The spectrum and eigenfunctions for one of these cases are obtained from the other one by a symmetry transformation.

\subsection{Complex factorizations}

The Hamiltonian (\ref{hv}) at the same time admits  a complex factorization.
As in the real case, here also there are two types of factorization due to a reflection symmetry of the potential involving both parameters:
\begin{equation}\label{cc}
\begin{array}{lll}
a)\qquad & \alpha\to i\beta,\quad & \beta\to -i\alpha
\\[2.ex]
b)\qquad & \alpha\to -i\beta,\quad & \beta\to i\alpha
\end{array}
\end{equation} 
From the first factorization (a),  we get the first order diferential operators $C_{\alpha,\beta}^{\pm}$ and the Hamiltonian  rewriten as:
\begin{equation}\label{cpm1}
C_{\alpha,\beta}^{\pm}=\mp\frac{d}{dx}+\frac{i\,\beta}{{\rm cosh}2x}-i(\alpha-i) {\rm tanh}2x, \qquad \mu_{\alpha}=(\alpha-i)^2
\end{equation}
\begin{equation}\label{h11c}
H_{\alpha,\beta}=C_{\alpha,\beta}^+ C_{\alpha,\beta}^- +\mu_{\alpha}
\end{equation}
where $\mu_{\alpha}$ has a complex value.

By means of the second  symmetry  
we get the second complex factorization
\begin{equation}\label{cpm2}
\tilde{C}_{\alpha,\beta}^{\pm}=\mp\frac{d}{dx}-\frac{i\,\beta}{{\rm cosh}2x}+i(\alpha+i) {\rm tanh}2x, \qquad \tilde{\mu}_{\alpha}=(\alpha+i)^2
\end{equation}
\begin{equation}\label{h12c}
H_{\alpha,\beta}=\tilde{C}_{\alpha,\beta}^+ \tilde{C}_{\alpha,\beta}^- +\tilde{\mu}_{\alpha}=C_{\alpha+2i,\beta}^- C_{\alpha+2i,\beta}^+ +\mu_{\alpha+2i}
\end{equation}

It is easy  to conclude from (\ref{cpm1}) that  the operators $C_{\alpha,\beta}^{\pm}$ are not adjoint each other: $(C_{\alpha,\beta}^{+})^{\dag}\neq C_{\alpha,\beta}^{-}$ but,
\begin{equation}\label{cpm}
(C_{\alpha,\beta}^{+})^{\dag} = (C_{\alpha,\beta}^{-})^*
\end{equation}
A relation that will be used to prove orthogonality relations later on.
Now, comparing (\ref{h11c}) and (\ref{h12c}), we are able to express the factorizations and intertwining relations as
\begin{figure}[ht]
  \centering
\includegraphics[width=0.4\textwidth]{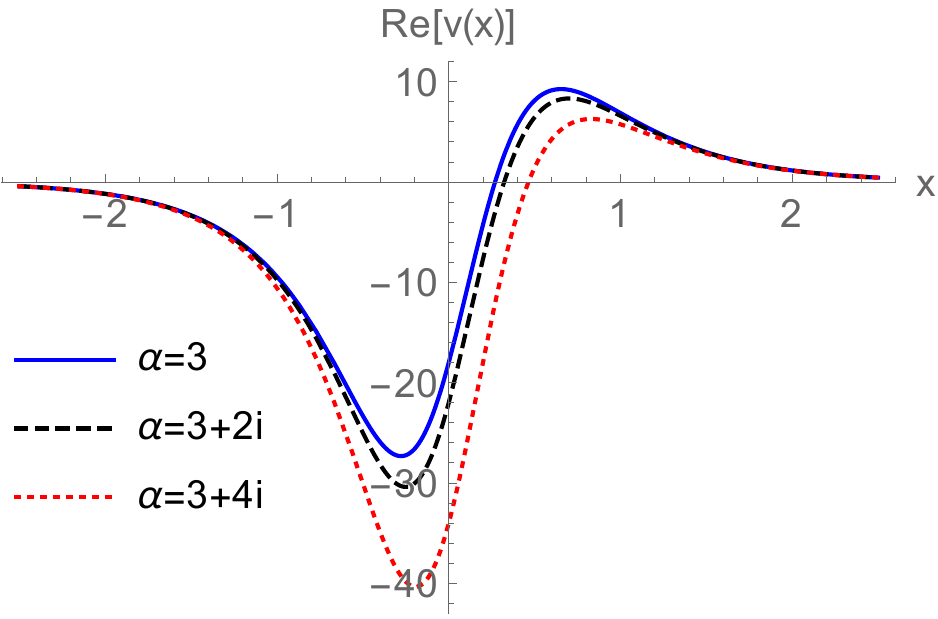}%
\hspace{1.5cm}%
  \includegraphics[width=0.4\textwidth]{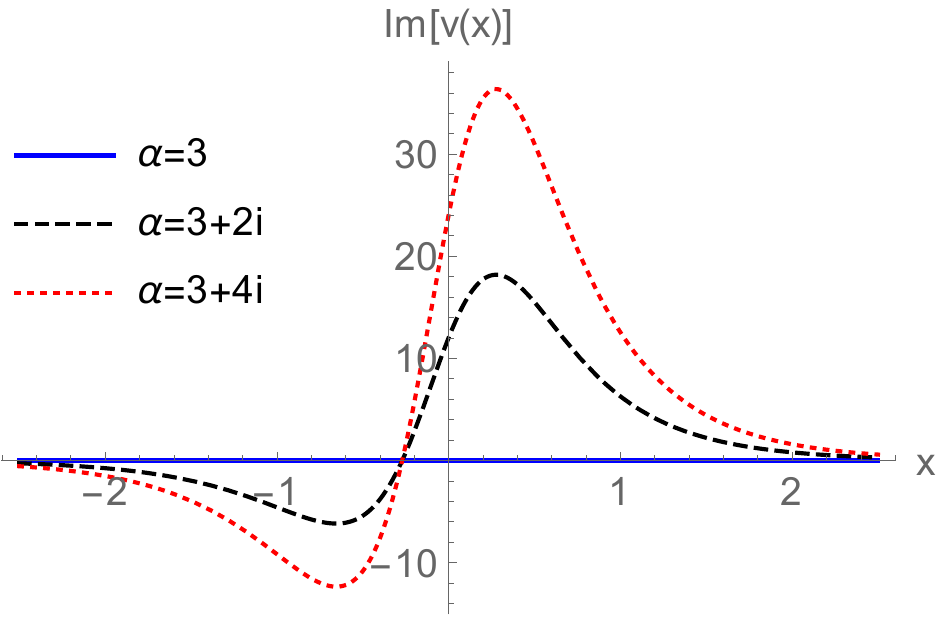}
\caption{\small
Plot of the real part (left) and the imaginary part (right) of the complex Scarf II potential for different values of parameters: $\beta=5.3, \alpha_1=3, \alpha_2=3+2i, \alpha_3=3+4i $. These are the complex potentials of the complex hierarchy corresponding $\beta=5.3$.
}
  \label{figuras3}
\end{figure}
\begin{equation}\label{h12hc}
H_{\alpha,\beta}=C_{\alpha,\beta}^+C_{\alpha,\beta}^- +{\mu}_{\alpha}=C_{\alpha+2i,\beta}^- C_{\alpha+2i,\beta}^+ +\mu_{\alpha+2i}
\end{equation}
\begin{equation}\label{int2}
C_{\alpha,\beta}^- H_{\alpha,\beta}=H_{\alpha-2i,\beta}C_{\alpha,\beta}^- ,\qquad C_{\alpha,\beta}^+ H_{\alpha-2i,\beta}=H_{\alpha,\beta}C_{\alpha,\beta}^+
\end{equation}
In this way, we arrive at the  hierarchy of complex Hamiltonians
$H_{\alpha+2i k,\beta}$, $k\in \mathbb Z$. {From Fig. \ref{figuras3}, it can be seen the real part (left) and the imaginary part (right)  of the complex Scarf II potential for different values of parameters. As in the real case, the parameter $\alpha$ does not affect the spectrum of the complex Scarf II Hamiltonians $H_{\alpha+2ik,\beta}$.}

\begin{figure}[ht]
  \centering
\includegraphics[width=0.325\textwidth]{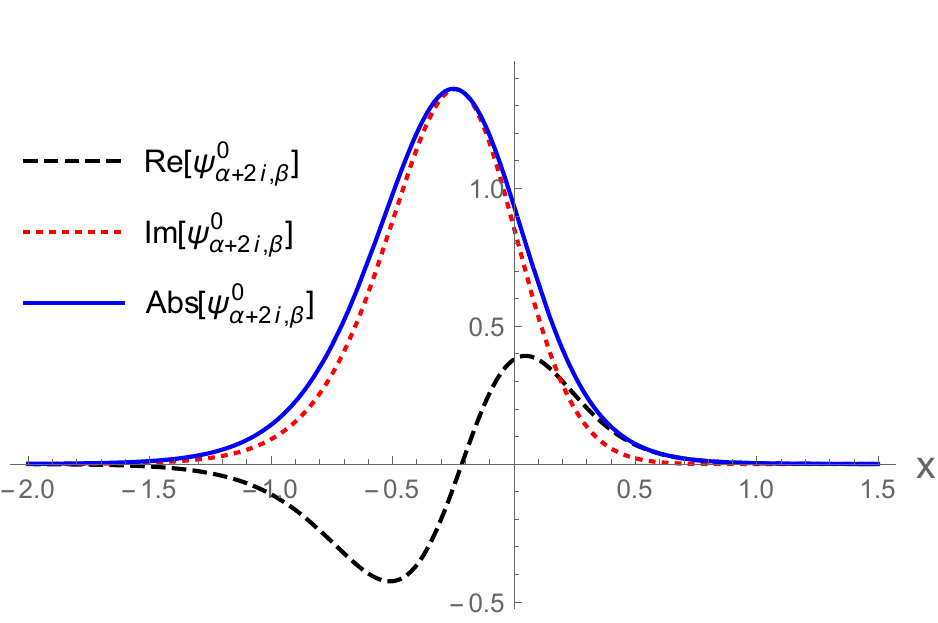}
  \includegraphics[width=0.325\textwidth]{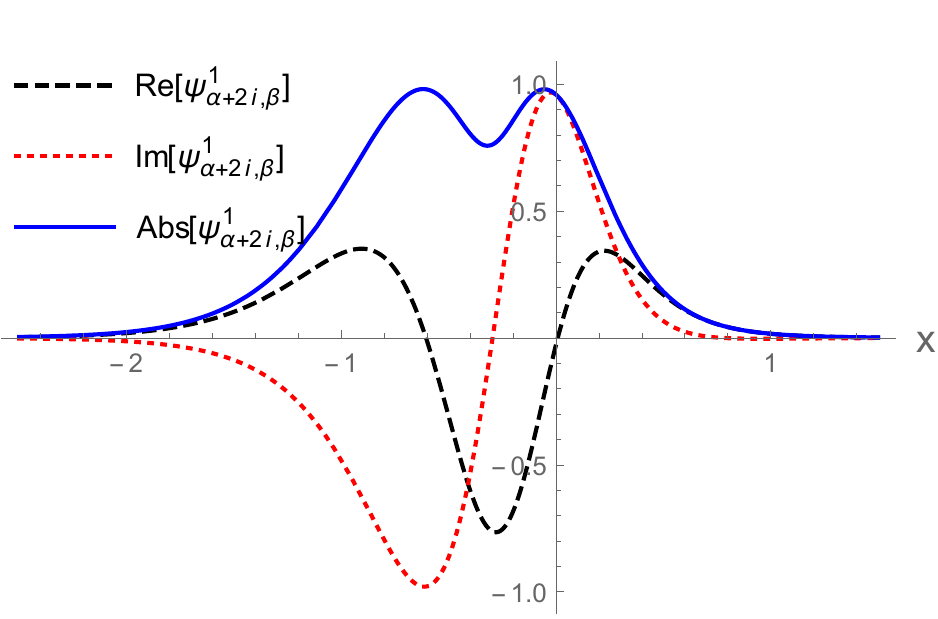}
  \includegraphics[width=0.325\textwidth]{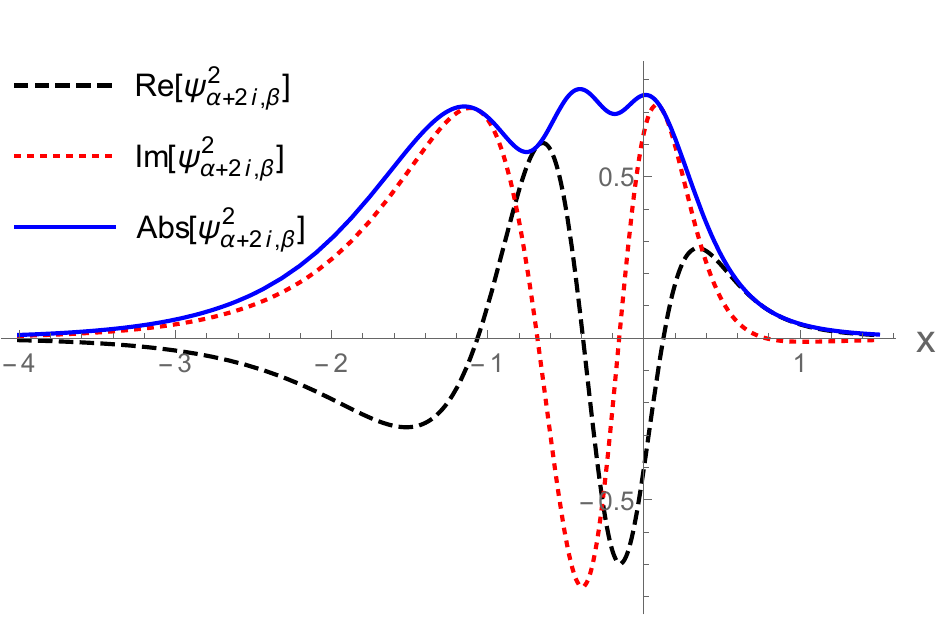}
\caption{\small
Plot of the solutions of the complex Scarf II potential for $\alpha=3+2i, \beta=6.8$  for  $n=0,1,2$.
}
  \label{figuras4}
\end{figure}

We try to find  a ground state $\phi_{\alpha,\beta}^0$ in the usual way by means of shift operators $C_{\alpha,\beta}^\pm$. So, we have the following equations to solve:
\begin{equation}\label{psi0cm}
C_{\alpha,\beta}^- \phi_{\alpha,\beta}^0=0\,,
\qquad
C_{\alpha+2i,\beta}^+ \phi_{\alpha,\beta}^0=0
\end{equation}
However, the solutions  obtained from (\ref{psi0cm}) are not square integrable. Thus, we conclude that there is no ground state energy solution annihilated by complex shift operators $C_{\alpha,\beta}^{\pm}$  and the supersymmetry is spontaneously broken \cite{CKS,Junker}. Nevertheless,   in order to find the eigenfunctions and eigenvalues of a  Hamiltonian in the complex hierarchy, for example $H_{\alpha-2i,\beta}$, we can use the solutions of  the real Hamiltonian $H_{\alpha,\beta}$.
Applying the shift (intertwining) operators $C_{\alpha,\beta}^\pm$ on these real solutions ($\psi_{\alpha,\beta}^n$), we get the bound state solutions of the complex Hamiltonian: 
\begin{equation}\label{complex1}
C_{\alpha,\beta}^-\psi_{\alpha,\beta}^n \propto \psi_{\alpha-2i,\beta}^n, \qquad C_{\alpha,\beta}^+\psi_{\alpha-2i,\beta}^n \propto \psi_{\alpha,\beta}^n
\end{equation}
where $n=0,1,2,\dots, n_{max}$ and the  maximum value of $n$  depends on the value of  $\beta$ as in the real case: $n_{max}=[\frac{\beta-1}{2}], \, \beta>1$. These complex solutions $\psi_{\alpha-2i,\beta}^n$ have the same real eigenvalue as the initial real solutions $\psi_{\alpha,\beta}^n$:
\begin{equation}\label{complex1}
E_{\alpha,\beta}^n =E_{\alpha-2i,\beta}^n
\end{equation}
We have checked that indeed these complex solutions are square integrable, which can be proved from the expression (\ref{cpm1}) of $C_{\alpha,\beta}^{\pm}$. Some of them are represented in Fig.~\ref{figuras4}. The same happens for the rest of complex Hamiltonians $H_{\alpha+2k i,\beta}$, for $k= 0,\pm1,\pm2,\dots$\,. They are isospectral to the real $H_{\alpha,\beta}$, although their solutions are complex. For example, $nth$ excited state solution $\psi_{\alpha-2ki,\beta}^n$, with $k\in\mathbb Z^+$, is found by acting with
 $C_{\alpha,\beta}^-$ as usual 
\begin{equation}\label{complex2}
C_{\alpha-(2k+2)i,\beta}^-\dots C_{\alpha,\beta}^-\psi_{\alpha,\beta}^n \propto \psi_{\alpha-2ki,\beta}^n\,,
\qquad E_{\alpha,\beta}^n =E_{\alpha-2k i,\beta}^{n}
\end{equation}

The complex solutions $\psi_{\alpha-2ki,\beta}^n$, $n=0,1\dots n_{\rm max}$ of the complex Hamiltonians
$H_{\alpha-2k i,\beta}$ are not orthogonal in the usual sense but they are
with respect to another type of product. Let us check, for example that the
two  eigenfunctions $\psi_{\alpha-2i,\beta}^{n_j}$, $n_j = n_1,n_2$ with
$n_1\neq n_2$ are orthogonal in the following sense.
Define the product by (see \cite{Znojil01,Bender04})
\begin{equation}\label{prod1}
\langle \psi_{\alpha-2i,\beta}^{n_1} , \psi_{\alpha-2i,\beta}^{n_2} \rangle:= \int_{-\infty}^{\infty} \psi_{\alpha-2i,\beta}^{n_1}\, \psi_{\alpha-2i,\beta}^{n_2}\, dx \propto 
\int_{-\infty}^{\infty} C^-_{\alpha,\beta} \psi_{\alpha,\beta}^{n_1}\,
C^-_{\alpha,\beta} \psi_{\alpha,\beta}^{n_2} \, dx
\end{equation}
Next, in this product, from the definition of $C^\pm_{\alpha,\beta}$
and the property (\ref{cpm}), the last integral is
\begin{equation}\label{prod2}
\langle \psi_{\alpha-2i,\beta}^{n_1} , \psi_{\alpha-2i,\beta}^{n_2} \rangle \propto 
\int_{-\infty}^{\infty} C^+_{\alpha,\beta}C^-_{\alpha,\beta} \psi_{\alpha,\beta}^{n_1}\,
 \psi_{\alpha,\beta}^{n_2} \, dx = 
\int_{-\infty}^{\infty}   \psi_{\alpha,\beta}^{n_1}\,
C^+_{\alpha,\beta}C^-_{\alpha,\beta} \psi_{\alpha,\beta}^{n_2}\, dx 
\end{equation}
Thus, taking into account (\ref{h11c}), we get 
\begin{equation}\label{prod3}
\langle \psi_{\alpha-2i,\beta}^{n_1} , \psi_{\alpha-2i,\beta}^{n_2} \rangle \propto 
(E_{\alpha,\beta}^{n_1}- \mu_\alpha)\int_{-\infty}^{\infty}  \psi_{\alpha,\beta}^{n_1}\,
\psi_{\alpha,\beta}^{n_2} \, dx  
=(E_{\alpha,\beta}^{n_2}- \mu_\alpha)
\int_{-\infty}^{\infty}   \psi_{\alpha,\beta}^{n_1}\,\psi_{\alpha,\beta}^{n_2}\, dx 
\end{equation}
From the formula (\ref{prod3}), we conclude that this equality is satisfied only if 
\begin{equation}\label{prod4}
\langle \psi_{\alpha-2i,\beta}^{n_1} , \psi_{\alpha-2i,\beta}^{n_2} \rangle =0 
\end{equation}
We have checked numerically, that indeed these complex eigenfunctions are orthogonal in this sense. As a consequence, the ``norm'' of 
$\psi_{\alpha-2i,\beta}^{n}:=C^-_{\alpha,\beta} \psi_{\alpha,\beta}^{n}$ is given by
\begin{equation}\label{prod5}
\langle C^-_{\alpha,\beta} \psi_{\alpha,\beta}^{n}\,,
C^-_{\alpha,\beta} \psi_{\alpha,\beta}^{n} \rangle =
(E_{\alpha,\beta}^{n}- \mu_\alpha) 
\langle \psi_{\alpha,\beta}^{n}\,,
\psi_{\alpha,\beta}^{n} \rangle \neq 0
\end{equation}

{We have obtained complex Hamiltonians $H_{\alpha+2i k,\beta}$ with real spectrum by means of complex factorization operators 
$C^\pm_{\alpha\pm 2k i,\beta}$ without taking care of PT symmetry. The conditions to get a real spectrum is based on two points: i)  The initial Hamiltonian $H_{\alpha,\beta}$ is real with real spectrum; ii) the complex operators $C^\pm_{\alpha\pm 2k i,\beta}$ do not annihilate physical square integrable states. In these conditions we have arrived to isospectral complex Hamiltonians.
The conclusion of this section is that associated to the real factorization of Scarf II systems, there is a set of complex factorizations leading to complex Hamiltonians.}

\section{Algebras of the operators}

\subsection{Algebra of real factorizations}
The real shift operators ${A}_{\alpha,\beta}^{\pm}$ together with an additional ``diagonal'' operator $A^0$, 
close the $su(1,1)$ Lie algebra \cite{Perelomov,Vilenkin}. In order to see this property, we   rewrite (\ref{h12h}) in the form:
\begin{equation}\label{h12hx}
 A_{\alpha,\beta+2}^- A_{\alpha,\beta+2}^+
 -
 A_{\alpha,\beta}^+A_{\alpha,\beta}^-
 =
-\mu_{\beta+2}+ {\mu}_{\beta}= 4\beta
\end{equation}
Next, let us define the natural operators (without indices) $A^\pm$ and $A^0$ acting on the eigenfunctions $\psi^n_{\alpha,\beta}$ of any Hamiltonian $H_{\alpha,\beta}$ in the form
\begin{equation}\label{aes}
A^-\psi_{\alpha,\beta}^n:=\frac12 A_{\alpha,\beta}^-\psi_{\alpha,\beta}^n\propto \psi_{\alpha,\beta-2}^{n-1},\quad
A^+\psi_{\alpha,\beta}^n := \frac12 A_{\alpha,\beta+2}^+\psi_{\alpha,\beta}^n\propto \psi_{\alpha,\beta+2}^{n+1},\quad
A^0 \psi_{\alpha,\beta}^n:=\frac12 \beta \psi_{\alpha,\beta}^n
\end{equation}
They satisfy the $su(1,1)$ algebra, where in the following expressions it is assumed that they are acting on 
an eigenfunction $\psi_{\alpha,\beta}^n$:
\begin{equation}\label{su11a}
[ A^-,A^+]= 2 A^0,\qquad [A^0, A^{\pm}]=\pm A^{\pm} 
\end{equation}
In this case, the Casimir operator is
\begin{equation}\label{cas}
{\cal C}_{\rm su(1,1)} = A^+A^- - A^0(A^0-1) = A^-A^+ - A^0(A^0+1)
\end{equation}
If we act ${\cal C}_{\rm su(1,1)}$ on the fundamental state (\ref{psi0}), $\psi_{\alpha,\beta}^0(x)$, we
get
\begin{equation}\label{cas2}
{\cal C}_{\rm su(1,1)} = -\frac{\beta}2 (\frac{\beta}2-1):= -\nu(\nu-1),\qquad \nu = \beta/2,\quad\beta >1
\end{equation}
This is a lowest bounded $\nu$--representation of $su(1,1)$ and the support space is spanned by the eigenfunctions   $\psi_{\alpha,\beta+2n}^n(x)$, $n=0,1,2,\dots$ Notice that according
to (\ref{h11}), and the definition (\ref{aes}),
\begin{equation}\label{h11b}
H \, \psi_{\alpha,\beta}^n:=H_{\alpha,\beta}\, \psi_{\alpha,\beta}^n=(4 A^+ A^- -(\beta-1)^2) \psi_{\alpha,\beta}^n
\end{equation}
Comparing with the Casimir operator (\ref{cas}) and taking into account the definition of $A^0$ in (\ref{aes})
we have 
\begin{equation}\label{cas2b}
{\cal C}_{\rm su(1,1)} =  \frac14 (H + 1),\quad\beta >1
\end{equation}
Therefore, the value of the Hamiltonian on the eigenfunctions $ \psi_{\alpha,\beta+2n}^n(x)$ is related to the value of the Casimir by (\ref{cas2b}) and it is given by (\ref{cas2}).

In the case $\tilde \beta<-1$,
if we apply the Casimir (\ref{cas}) on the fundamental state (\ref{psi0b}), 
$\tilde\psi_{\tilde\alpha,\tilde\beta}^0(x)$ we find
\begin{equation}\label{cas3}
{\cal C}_{\rm su(1,1)} = -\frac{\tilde\beta}2 (\frac{\tilde \beta}2+1)=
-\frac{\beta}2 (\frac{ \beta}2-1) := -\nu(\nu-1),\qquad \nu = -\tilde\beta/2,\quad\tilde\beta <-1
\end{equation}
This is an upper bounded $su(1,1)$ representation with the same Casimir eigenvalue as the previous lowest representation. 

In conclusion, positive values of $\beta>1$ give rise to a lowest bounded representation while the opposite values $\tilde\beta = -\beta <-1$ lead to upper
bounded $su(1,1)$ representations with the same Casimir and Hamiltonian eigenvalues.
\subsection{Algebra of complex factorizations}

For the complex case, the complex shift operators also close an algebra together with an additional diagonal operator. In order to see this property, we have to define again the  operators $C^\pm,\, C^0$ having in mind (\ref{complex1}):
\begin{equation}\label{ces}
C^-\psi_{\alpha,\beta}^n :=\frac12 C_{\alpha,\beta}^-\psi_{\alpha,\beta}^n\propto \psi_{\alpha-2i,\beta}^{n},\quad
C^+\psi_{\alpha,\beta}^n := \frac12 C_{\alpha,\beta+2}^+\psi_{\alpha,\beta}^n\propto \psi_{\alpha+2i,\beta+2}^n,\quad
C^0 \psi_{\alpha,\beta}^n:=\frac12 \alpha \psi_{\alpha,\beta}^n
\end{equation}
They satisfy the following algebra (which will also correspond to  $su(1,1)$):
\begin{equation}\label{su11b}
[{ C}^-,{ C}^+]=2 i { C^0},\qquad [{ C^0},{ C}^{\pm}]=\pm \, i { C}^{\pm} 
\end{equation}
These are not the standard commutation relations of $su(1,1)$ \cite{Perelomov,Vilenkin,Olmo99}. However, in fact, they can be identified with the Lie algebra $su(1,1)$ in another basis. In the case of real factorizations
the commutators (\ref{su11a}) correspond to the case where the diagonal
operator $A^0$ represents a compact generator (of trigonometric rotations) while in the case of complex factorizations the commutators (\ref{su11b}) apply when the diagonal operator $C^0$ represents a noncompact operator (generating hyperbolic transformations).

The Casimir operator of the algebra (\ref{su11b}) 
 is
\begin{equation}\label{casxx}
{\cal C}_{\rm su(1,1)} = C^+C^- + C^0(C^0-i) 
\end{equation}
then, from (\ref{h12hc}) we get
\begin{equation}\label{h12hcc}
{\cal C}_{\rm su(1,1)}=\frac14(H +1) 
\end{equation}
which is the same value as in the real $su(1,1)$ algebra (\ref{cas2b}).

\subsection{Complete algebra of Scarf II factorizations}

\begin{figure}[h]
  \centering
\includegraphics[width=1\textwidth]{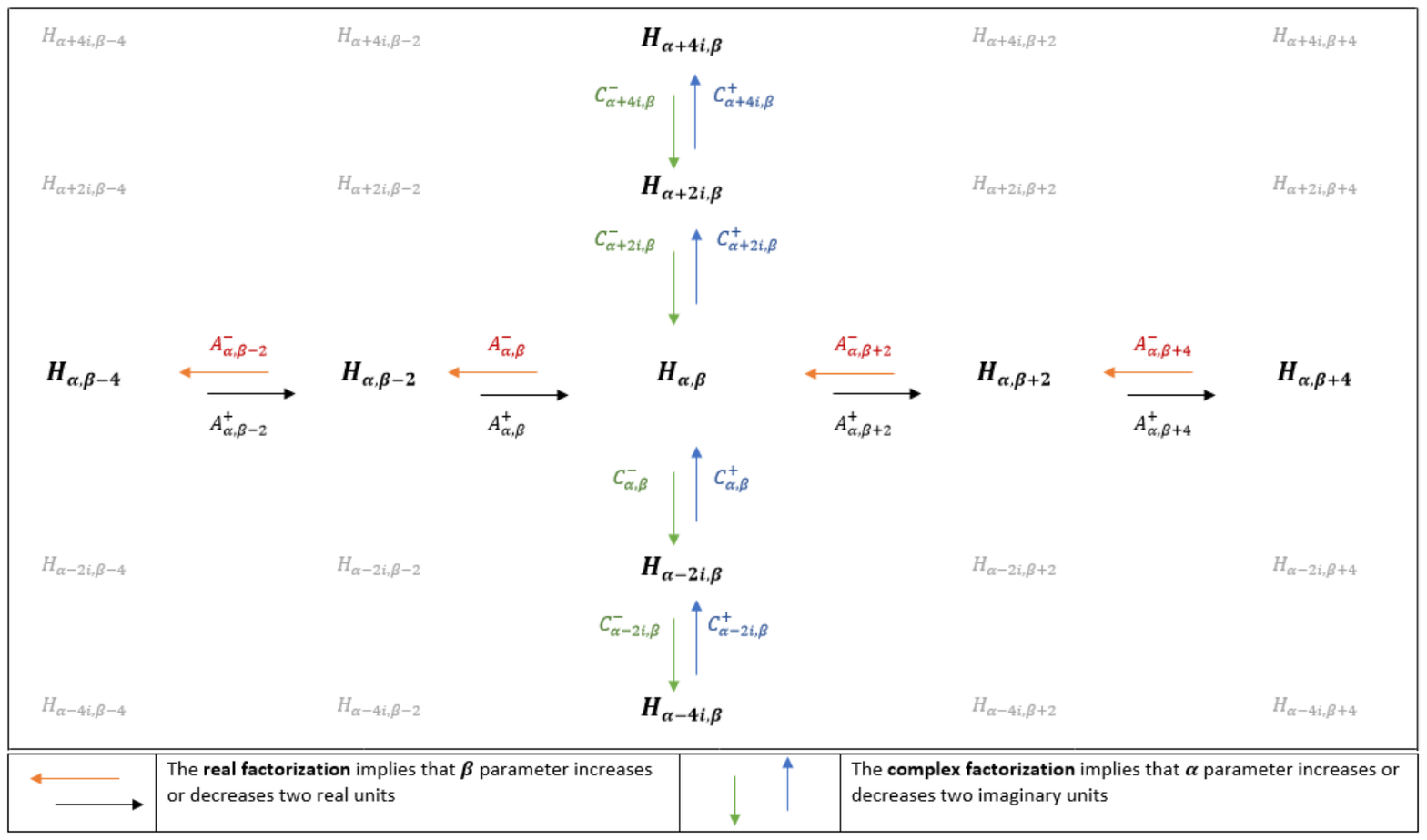}%
\vspace{-2cm}
\caption{\small
Schematic drawing of hierarchies of the real and the complex Scarf II potential Hamiltonians.
}
  \label{figuras5}
\end{figure}

At the same time, it can be easily checked that the real shift operators $A^{\pm},A^0$ and the complex shift operators $C^{\pm},C^0$ commute with each other,
\[
[C^{i}, A^{j}]=0\,,\qquad i,j= \pm,\ 0
\]
when they act on any eigenfunction $\psi_{\alpha+2ki,\beta+2m}^n$ of the hierarchy, and therefore the total shift (or potential) algebra for the double hierarchy is $su(1,1)\oplus su(1,1)$. As mentioned above, the factorization operators ($A^\pm$ and $C^\pm$) do not correspond to the same basis of the $su(1,1)$ algebra,
this is the reason why one pair ($A^\pm$) produces real and the other ($C^\pm$) complex factorizations. An important
point is that their corresponding representations have the same Casimir eigenvalue. 
{Other shape invariant systems like the familiar trigonometric P\"oschl-Teller potential  have also double factorizations, but they are real.
In that case the Lie algebra is a direct sum of two $su(2)$ algebras: $su(2)\oplus su(2)$. For more details the reader may consult \cite{Negro09,Negro12}. There is a two dimensional lattice associated to the double factorization in a similar way as Fig.~\ref{figuras5}.}

{The relation of factorizations with Lie algebras, can be understood, in some cases, from higher dimensional free systems having Lie symmetries. After separation of variables we obtain reduced systems that constitute factorizable effective Hamiltonians. The factorization operators come from the initial symmetries (see for instance \cite{Olmo99}).}

{We can appreciate clearly the action of real $A^{\pm}$ and complex  $C^{\pm}$  operators on the eigenfuncions of the Hamiltonians for the whole  hierarchy in Fig.~\ref{figuras5}. Each point of the lattice represent a Hamiltonian, but also can be interpreted as its eigenfunctions. The eigenfunctions of Hamiltonians on a horizontal row of the lattice connected by $A^{\pm}$ have the same energy. The complex Hamiltonians corresponding to the points of a vertical column are  connected by $C^{\pm}$ and they are isospectral to the real Hamiltonian in that column, $H_{\alpha,\beta}$. } 
The general Hamiltonians  and  the potentials in this hierarchy are given by
\begin{equation}\label{hg}
H_{\alpha+2k i,\beta+2m}(x)=-\frac{d^2}{dx^2}+V_{\alpha+2k i,\beta+2m}(x)
\end{equation}
\begin{equation}\label{vg}
V_{\alpha+2k i,\beta+2m}(x)=\frac{({\alpha+2k i})^2-{(\beta+2m)}^2+1+2 (\alpha+2k i)(\beta+2m) {\rm sinh}{2x}}{({\rm cosh}2x)^2}
\end{equation}
where $m, k \in \mathbb Z$.
These potentials can be written in a explicit complex form as:
\begin{equation}\label{vgim2}
V_{\alpha+2k i,\beta+2m}(x)=\frac{{\alpha}^2-(2k)^2+(\beta+2m)^2+1+2\alpha(\beta+2m){\rm sinh}{2x}}{({\rm cosh}2x)^2}
\\[2.ex]
+i\left(\frac{4 \alpha k+4k(\beta+2m){\rm sinh}{2x}}{({\rm cosh}2x)^2}\right)
\end{equation}
From here, we see that only if $k=0$, the potentials are real. This property is also displayed in Fig.~\ref{figuras5}, where each vertical or horizontal lines represent other (complex) one--parameter hierarchies. We notice that real and complex part of (\ref{vgim2}) have the form of real Scarf II potentials. { We can also see directly from (\ref{vgim2}) that this is not a PT-invariant potential
(unless for the case where the coefficient $2\alpha(\beta+2m)$ be zero).} If $\alpha=0$, this potential corresponds to complexified Scarf II potential given in the literature \cite{Ahmed01}. So, the potential given by (\ref{vgim2}) is a general form of Scarf II potential which includes  both real and complexified form.

We note that for the general case of the potential given in (\ref{v}), the change of the parameters $\alpha$ and $\beta$ instead of being $\Delta\alpha=\pm2i$ and $\Delta\beta=\pm2$, will be $\Delta\alpha=\pm \gamma i$ and $\Delta\beta=\pm\gamma $.  In this case, this potential have the same type of factorizations properties. 

\section{Conclusions}
In this work, we have described a kind of special isospectral factorizations of real Hamiltonians corresponding to Scarf II  potential which arises as a companion of the standard real factorization. This second factorizations are complex: the parameters of their potentials in this hierarchy vary in imaginary values. We have remarked that in this respect Scarf II potential is  unique  among the shape invariant factorizable potentials (for example, given in table 4 of \cite{CKS}). The obtained  complex Hamiltonian hierarchies have a real spectrum; the supersymmetry is spontaneously broken and the potentials are not PT-symmetric. 
The eigenfunctions and spectrum of all the complex potentials are derived from the real potential hierarchy. These relations are illustrated in Fig.~\ref{figuras5}. {This type of complex potentials are different from the wide family of complexified Scarf II potential introduced before, although it might be included in some of them. Complexified Scarf II potentials have applied in different ways to optical systems, working with the refractive index or with absortion properties, for example as shown in \cite{Kapoor22}.  We hope that the complex Scarf II potentials introduced here also can be applied in the future in the problems of optical systems.}

We point out that the total shift (or potential) algebra for this problem is the direct sum $su(1,1)\oplus su(1,1)$. This is due to the fact that the complex $C^\pm$ and real $A^\pm$ shift operators commute. This leads to a global two--dimensional lattice of parameters $(\alpha+2k i,\beta+2m)$, where $m, k \in \mathbb Z$, where each  point represent a Hamiltonian and a horizontal (or vertical) line represents a Hamiltonian hierarchy. Any two Hamiltonians in this global lattice can be linked by a product of complex and real shift operators.

\section*{Acknowledgements}

This work was  supported by MCIN, Spain with funding from the European Union
NexGenerationEU (PRTRC17.I1) and the Consejer\'ia de Educaci\'on de la Junta de Castilla y Le\'on, Spain, Project QCAYLE and the MCIN Project PID2020-113406GB-I00 of Spain. L. Acevedo acknowledges financial support from Doctorate ProgramFunds No. UVa2023 and Banco Santander. Y. C. Acar acknowledges TUBITAK 2210/A National MSc Scholarship Program. We thank Dr. Javier Negro for helpful comments and suggestions.
\\
\\
Data Availability Statement: No Data associated in the manuscript.

\bigskip

\bibliographystyle{plain}

\end{document}